\documentclass[aps,pre,reprint,groupedaddress,showpacs]{revtex4-1}
\usepackage[english]{babel}
\usepackage[latin2]{inputenc}
\usepackage{amsmath}
\usepackage{verbatim}

\begin{document}
\title{Collectivity in diffusion of colloidal particles: from effective interactions to spatially correlated noise}

\author{M. Majka}
\email{maciej.majka@uj.edu.pl}
\affiliation{Marian Smoluchowski Institute of Physics, Jagiellonian University, ul. prof. Stanis\l{}awa \L{}ojasiewicza 11, 30-348 Krak\'{o}w Poland}
\author{P. F. G\'{o}ra}
\affiliation{Marian Smoluchowski Institute of Physics, Jagiellonian University, ul. prof. Stanis\l{}awa \L{}ojasiewicza 11, 30-348 Krak\'{o}w Poland}

\begin{abstract}
The collectivity in the simultaneous diffusion of many particles, i.e. the interdependence of stochastic forces affecting different particles in the same solution, is a largely overlooked phenomenon with no well-established theory. Recently, we have proposed a novel type of thermodynamically consistent Langevin dynamics driven by the Spatially Correlated Noise (SCN) that can contribute to the understanding of this problem. This model draws a link between the theory of effective interactions in binary colloidal mixtures and the properties of SCN. In the current article we review this model from the perspective of collective diffusion and generalize it to the case of multiple ($N>2$) particles. Since our theory of SCN-driven Langevin dynamics has certain issues that could not be resolved within its framework, in this article we also provide another approach to the problem of collectivity. We discuss the multi-particle Mori-Zwanzig model, which is fully microscopically consistent. Indeed, we show that this model supplies many information, complementary to the SCN-based approach, e.g. it predicts the deterministic dynamics of the relative distance between the particles, it provides the approximation for non-equilibrium effective interactions and predicts the collective subdiffusion of tracers in group. These results provide the short-range, inertial limit of the earlier model and agree with its predictions under some general conditions. In this article we also review the origin of SCN and its consequences for the variety of physical systems, with emphasis on the colloids. 
\end{abstract}


\maketitle

\section{Introduction}

The classical picture of a single-particle diffusion, dating back to Einstein and Smoluchowski, describes the observed particle (a tracer) as constantly bombarded by the much smaller particles of environment \cite{bib:smoluchowski1, bib:smoluchowski2}. This results in the stochastic driving, which is accompanied by the Stokesian, hydrodynamic friction that dissipates the energy of the tracer. In the following decades the major modification to this picture came from Mori and Zwanzig who provided a formal link between the microscopic interactions in the system and the Langevin dynamics \cite{bib:zwanzig}. They have shown that the stochastic force is time-correlated and must be accompanied by the friction with memory kernel. This establishes the formalism of Generalized Langevin Equations \cite{bib:zwanzig, bib:hanggi}. Nowadays, the time-correlated noise and its consequences, most notably the sub-diffusive processes \cite{bib:kou, bib:goychuk, bib:metzler}, are thoroughly researched phenomena with multiple applications (e.g. \cite{bib:subdiff1, bib:subdiff2}). 

While the concept of the correlations in noise can be extended to the spatial domain, this aspect of stochastic dynamics remains widely unrecognized. In Sec. \ref{sec:SCNphenomena} we provide the summary of phenomena related to, what we call, Spatially Correlated Noise (SCN). Let us just mention now that SCN manifests in such problems as the self-assembly and its non-equilibrium dynamics \cite{bib:majka1} and  these problems are increasingly important in many branches of physics.

Few attempts have been aimed at understanding the role of SCN in the molecular systems and even fewer try to explain its origin. In our recent article (Ref.\cite{bib:majka1}), we have employed the binary mixture theory to show that the spatial correlations are inherently present in the particle-thermal bath interactions. We have also introduced a new type of the thermodynamically consistent SCN-driven Langevin dynamics \cite{bib:majka1}. This new formalism has been developed for two particles. In this article we extend it to the multi-particle case. We also introduce the idea of collectivity in diffusion, i.e. the notion that the friction and stochastic force affecting different particles are not independent. Finally, we introduce another approach to the spatially correlated behavior, which is the multi-particle Mori-Zwanzig model. We will show that this model proves complementary to the SCN-based theory and can help us resolve some of its ambiguities.

We will now give a short overview of Ref. \cite{bib:majka1}, which sets the context for this current work. Let us assume that microscopically the system consists of $N$ tracer particles and $\tilde N$ particles of thermal bath. The total microscopic Hamiltonian of this system reads:
\begin{equation}
\mathcal{H}=\mathcal{H}_{tt}+\mathcal{H}_{tb}+\mathcal{H}_{bb} \label{eq:H1}
\end{equation}
where $\mathcal{H}_{tt}$ contains the microscopic tracer-tracer interactions, $\mathcal{H}_{tb}$ is the tracer-bath interaction and $\mathcal{H}_{bb}$ is the Hamiltonian of thermal bath, including its internal interactions. More specifically:
\begin{equation}
\mathcal{H}_{tt}=\sum_i^N\frac{P^2_i}{2M}+\sum_{i>j}^NU_0(x_i-x_j)
\end{equation}
\begin{equation}
\mathcal{H}_{tb}=\sum_i^N\sum_j^{\tilde N}V(x_i-q_j)
\end{equation}
\begin{equation}
\mathcal{H}_{bb}=\sum_i^{\tilde N}\frac{p_i^2}{2m}+\sum_{i>j}^{\tilde N}v(q_i-q_j) \label{eq:H2}
\end{equation}
where $(x_i,P_i)$ and $(q_i,p_i)$ are the phase-space coordinates of the tracers and environment particles, respectively. In this setting, we can determine the effective interactions, i.e. the interactions that in equilibrium are equivalent to the influence of the thermal bath on tracers \cite{bib:majka2, bib:lekkerkerker}. These interactions read:
\begin{equation}
\begin{split}
&\mathcal{H}_{eff}=\sum_{i>j}^N U_{eff}(x_i-x_j)=\\
&=-\frac{1}{\beta}\ln\left(\int\{dp dq\}\exp\left(-\beta(\mathcal{H}_{Tb}+\mathcal{H}_{bb}) \right) \right)
\end{split} \label{eq:Ueff}
\end{equation}
where $\{dpdg\}=\prod_{i}^{\tilde N}dp_idq_i$ and $\beta=(k_BT)^{-1}$ is the temperature factor. We can define now the deterministic tracer-bath coupling force as:
\begin{equation}
\xi(x_i)=-\sum_j^{\tilde N}\partial_{x_i}V(x_i-q_j)
\end{equation}
We have shown that the covariance of this coupling force reads \cite{bib:majka1}:
\begin{equation}
<\xi(0)\xi(r)>=<\sum_{\substack{j\neq i,\\k\neq i}}^N F_{eff}(x_i-x_j)F_{eff}(x_i-x_k+r)> \label{eq:corr}
\end{equation}
The average is taken with respect to the Boltzmann distribution:
\begin{equation}
\begin{split}
&P_B(x_1,...,x_N)=\mathcal{N}^{-1}\exp\left(-\beta \sum_{i>j}U(x_i-x_j) \right) 
\end{split} \label{eq:P_Boltz}
\end{equation}
where $U(r)=U_0(r)+U_{eff}(r)$ and $\mathcal{N}$ is the normalization constant. Then, the spatial correlation function of the tracer-bath coupling force reads:
\begin{equation}
h(r)=<\xi(0)\xi(r)>/<\xi^2(x_i)>
\end{equation}
Thus, these correlations proves to be directly related to the correlations of effective forces. This result provides the microscopic foundation for SCN which relates it to the wide class of soft matter phenomena. Further, we introduce the Langevin dynamics of two tracers driven by the Gaussian SCN, in which the noise correlation function is given by $h(r)$. We have found that the thermodynamic consistency (i.e. recovering the Boltzmann distribution \eqref{eq:P_Boltz} in the steady state) requires that the friction coefficient for the relative variable $r=x_2-x_1$ is spatially variant. More precisely, we have derived the following over-damped equation of motion:
\begin{equation}
K(r)\dot r=2F(r)+\sqrt{2}\sigma g_-(r)\eta \label{eq:SCN_Lan}
\end{equation}
where $F(r)=F_0(r)+F_{eff}(r)$, $g_-(r)=\sqrt{1-h(r)}$, $\eta$ is the uncorrelated Gaussian noise, $h(r)$ is the correlation function given by \eqref{eq:corr}, $\sigma$ is the noise amplitude and:
\begin{equation}
K(r)=\frac{g_{-}(r)e^{-\beta U(r)}}{\frac{1}{\gamma}-\frac{2}{\sigma^2}\int_r^{+\infty}dr'\frac{F(r')}{g(r')}e^{-\beta U(r')}} \label{eq:K}
\end{equation}
is the Spatially Variant Friction Coefficient (SVFC). The application of this SCN-driven Langevin dynamics to the system of two charged spheres and counter-ions leads to several non-equilibrium effects, e.g. the emergence of the friction-less regime and the transient attraction effect \cite{bib:majka1}. 

These results settle the technical core of our theory and in this article we want propose its generalizations as well as discuss some of its additional aspects. First, in Sec. \ref{sec:SCNphenomena} we review numerous contexts in which SCN is encountered, speculating on the applicability of our theory. In Sec. \ref{sec:collectivity} we explain the notion of collectivity in diffusion, which is the major consequence of Ref. \cite{bib:majka1} and in Sec. \ref{sec:multiparticle} we extend our formalism to the systems with multiple ($N>2$) tracers. Further, we will confront our theory with another approach, which is the multi-particle Mori-Zwanzig model. In Ref. \cite{bib:majka1}, we have shown that SVFC tends to zero as the particles get close to each other. This behavior arise in the over-damped theory, which indicates its breakdown. Thus, the insight from the inertial dynamics is desired and we propose the multi-tracer Mori-Zwanzig model as a solution to this problem. The Mori-Zwanzig model is also the simplest, microscopically consistent approach to the collectivity in diffusion and we show that it provides the complementary perspective on our SCN-driven Langevin dynamics. However, we will also show that the Mori-Zwanzig approach is severely limited in its predictions and, in fact, it can only serve as an auxiliary short-distance approximation. The Mori-Zwanzig model is solved in Section \ref{sec:2particleMZmodel} and the comparison to SCN-driven dynamics is given in Section \ref{sec:comparison}.

\section{SCN in molecular systems} \label{sec:SCNphenomena}
SCN is a random disturbance statistically characterized by a certain correlation length-scale, i.e. the stochastic forces 'felt' by two nearby particles are similar in terms of their direction and amplitude, though the pattern of forces at a greater length-scale is completely random. Such pattern can also randomly evolve in time and if its memory is short, this system might be though of as purely SCN-driven. 

One prominent class of phenomena affected by the spatially correlated behavior are the dense molecular systems, especially in the glassy state \cite{bib:glass0}. It has been shown via both simulations and experiments that the velocities of particles in the glassy regime are spatially correlated \cite{bib:glass1,bib:glass2}, the rearrangement of molecules is highly cooperative \cite{bib:glass3} and the clusters of particles moving together are also observed \cite{bib:glass4}. These features strongly resemble SCN. Yet another discipline in which SCN occurs is the intracellular biophysics, e.g. it has been shown that the flows of cytoplasm inside a cell are spatially correlated \cite{bib:biophys1}. Recent research has proven that the most of the diffusive transport in cytoplasm results from the constant stirring by molecular motors \cite{bib:biophys2} and the other active components \cite{bib:biophys3}. Since the random flows caused by stirring are spatially extensive, the intracellular diffusion should be considered as being SCN-driven. This is partially related to the field of active matter in which the self-propelling particles are also known to produce SCN, e.g. in the interaction between the polymer chain and active swimmers \cite{bib:shin} or when active particles form clusters \cite{bib:active}. The spatially correlated behavior is also encountered in fluids, e.g. in the sedimentation experiments (as a result of hydrodynamic interactions, e.g. \cite{bib:sedimentation}) or in the turbulent flows \cite{bib:turbulent}.

While the systems we mention span a wide range of physical phenomena they share a common approach in modeling, which is the Langevin dynamics. However, while the spatial correlations seem to play the crucial role in these systems, the ordinary Langevin dynamics neglects these correlations completely. Thus, the understanding of these systems might be improved with the perspective of SCN. However, the SCN-driven dynamics, which we discuss here is based on the equilibrium correlation function, so it is applicable in the close to equilibrium regime only. This means that the far-from equilibrium systems, especially the flows and active particles, lie beyond its current scope. However, the general observation that the SCN requires SVFC should a specific stationary distribution be recovered, might remain valid for these systems. The adaptation of our approach would require identifying the non-equilibrium distributions. The glassy systems are a more promising field, since they are the limiting case of the equilibrium colloidal systems, for which our theory is applicable.

The two-component (binary) colloids are the class of systems for which our SCN-driven Langevin dynamics has been derived. In these systems the self-organization is induced by the effective interactions \cite{bib:majka2,bib:lekkerkerker} as given by \eqref{eq:Ueff}. The relation \eqref{eq:corr} shows that the presence of effective interactions entails the presence of spatial correlations in the tracer-thermal bath interactions. This suggest that the self-assembly might be dynamically perceived as SCN-driven. The simplest example of this phenomenon is e.g. the clustering of colloidal spheres due to the presence of smaller particles \cite{bib:yodh}. In this example, described on the statistical level by the renowned Asakura-Oosawa model \cite{bib:asakura}, the bigger spheres do not interact in the long-range fashion, but they are kept together by the interaction with environment. Although for the hard-spheres one might explain this effect by a simple imbalance in the number of molecular collisions \cite{bib:lekkerkerker}, there are more subtle aspects of this phenomenon e.g. the multiple energy minima of this interaction in a dense environment \cite{bib:yodh} or the long-range 'attraction-through-repulsion'/'repulsion-through-attraction' effects for screened-charged spheres \cite{bib:majka2, bib:louise} that suggest the cooperativeness in the influence of environment. In fact, in the light of our SCN-driven Langevin dynamics, any type of self-organization driven by the effective interactions (so e.g. the separation of polymer blends \cite{bib:majka3}, like-charge attraction \cite{bib:likecharge} etc.) can be interpreted as the manifestation of SCN. We should also mention that many of the biological self-assembly mechanisms (e.g. \cite{bib:marenduzzo, bib:chromosome1, bib:chromosome2}) belong to this category, since the molecular crowding and excluded volume effects are important factors in the intracellular environment. However, we should mention that we are not aware of any direct experimental observations of SCN in colloidal systems. The reason is that the most of experimental techniques can directly measure the positions of particles, but deducing the random forces usually requires one to pre-assume some microscopic model. Thus, the spatial correlations are usually associated with the deterministic rather than stochastic forces. 

The  phenomena qualitatively similar to the colloidal self-assembly are also encountered in the plasma physics. The electro-magnetic field is \emph{per se} the spatially extensive and fluctuating entity and a few attempts have been made to incorporate this fact into the theoretical description of the diffusion in plasma (e.g. \cite{bib:plasma1,bib:plasma2}). However, these considerations address solely the diffusion of a single particle. On the other hand the two-component plasma can behave similarly to the binary colloids and the analogous separation phenomena emerge in both simulations and experiments \cite{bib:plasma3,bib:plasma4}. Some theoretical attempts (e.g. \cite{bib:plasma5}) describe these effects from the perspective of effective interactions.

The given examples show that SCN emerges in many different branches of physics. Despite this fact, the literature regarding the influence of SCN on molecular systems is rather scarce. One example is the research on the SCN-driven Single File Diffusion, i.e. the effectively one-dimensional diffusion of impenetrable particles in narrow pores, which leads to the sub-diffusive dynamics \cite{bib:SFD}. Another research utilizes  SCN as a stimulus for a neural network \cite{bib:neuron}, concluding that it is a crucial factor in obtaining the synchronized output. Yet another research has been carried out by the current authors, who have analyzed the influence of SCN on a model polymeric chain. We have found that SCN can cause the spontaneous unfolding effect and the synchronization of monomers motion \cite{bib:majka4, bib:majka5}, possibly leading to the non-Gaussian chain statistics \cite{bib:majka3}. This last idea finds a qualitative confirmation in the research on the influence of active swimmers on a polymer chain by Shin et al. \cite{bib:shin}.

Finally, from a purely theoretical perspective, a distinct class of stochastic differential equations with multiplicative noise can be related to the SCN-driven system. This comes from the properties of the correlated Gaussian noise. Let $\xi(x_i)$ be the Gaussian random variable that satisfies:
\begin{equation}
\begin{gathered}
<\xi(x_i)>=0\\
<\xi^2(x_i)>=\sigma^2\\
 <\xi(r)\xi(0)>/<\xi^2(x_i)>=h(r)
\end{gathered}
\end{equation}
where $h(r)$ is now some spatial correlation function. Knowing that the sum of Gaussian variables is also a Gaussian variable, one can write \cite{bib:majka1}:
\begin{equation}
\xi(x_1)\pm\xi(x_1+r)=\sqrt{2}\sigma\sqrt{1\pm h(r)}\eta=\sqrt{2}\sigma g_{\pm}(r)\eta \label{eq:trick}
\end{equation}
where $\eta$ is now the uncorrelated Gaussian noise. This shows that the linear combination of the additive, spatially correlated noise terms can be translated into the one, uncorrelated, but multiplicative noise term. What follows, any stochastic differential equation of the form
\begin{equation*}
\gamma \dot r=2F(r)+\sqrt{2}\sigma g_{-}(r)\eta
\end{equation*}
where $F(r)$ is an odd function, can be turned into the over-damped dynamics of two particles driven by SCN. Namely, given that $r=x_2-x_1$, their equations of motion read: 
\begin{equation*}
\gamma \dot x_i=F(x_i-x_j)+\xi(x_i)
\end{equation*}
where $<\xi(x_i)\xi(x_j)>/\sigma^2=1-g_{-}^2(r)$. From this perspective Ref. e.g. \cite{bib:denisov, bib:gora} can be considered as the analysis of the over-damped oscillator driven by SCN with $<\xi(x_i)\xi(x_j)>/\sigma^2=1-r^{2\alpha}$. It is shown there that such SCN induces the localization of particles and the probability distribution for $r$ is discontinuous in $\alpha$. 

Unfortunately, the disadvantage of the existing research on the SCN-driven systems is that it makes use of the constant friction coefficient. The crucial observation from our theory is that one cannot achieve the thermodynamic consistency without SVFC given by \eqref{eq:K}\cite{bib:majka1}. This means that none of these results can be related to the equilibrium conditions in a simple way. A similar observation regarding SCN appears independently in Ref. \cite{bib:nardini}. The notion of SVFC itself is also not entirely new and it can be encountered in the research on the diffusion in \emph{viscosity landscapes}, e.g. in \cite{bib:viscosity}. Interestingly, in this context the authors intuitively conclude that the system must be driven by the multiplicative noise. This suggests a weak analogy between their approach and our SCN-driven dynamics, i.e. one could interpret our SVFC as the diffusion in some effective viscosity landscape. In this context, it is particularly intriguing, whether the subdiffusive behavior predicted in \cite{bib:viscosity} could emerge in our approach. Unfortunately, as our equations of motion are highly nonlinear and the dependence between $K(r)$, $F(r)$ and $g(r)$ is strictly fixed (unlike in Ref. \cite{bib:viscosity}), finding an exactly solvable toy-model for our theory is still an open problem.

Finally, we should also mention that systems with SCN are often subjected to some significant temporal correlations. Unfortunately, the problem of spatio-temporal dynamics is extremely challenging and reaches beyond the scope of this paper. In this context, our SCN-based approach should be perceived as a step towards the ultimate spatio-temporal solution. 

\section{Collectivity in diffusion}\label{sec:collectivity}
Let us now discuss a major consequence of relation \eqref{eq:corr}, which is the collectivity in diffusion. When we describe the diffusion of the group of tracers driven by the common thermal bath this resembles the situation of the binary mixture. In other words, we divide the entire system into two species of particles (tracers and the particles of therm bath) and try to replace one of them with some effective impact on the other. The difference is that in soft matter we are usually interested in the stationary behavior of the system and in diffusion we ask for the dynamics. Nevertheless, the traced-out species of particles is always the source of effective forces and these fact should be reflected by the properties of thermal bath.

First, let us discuss the most basic approach. The mean value of any observable $O(\{x\})$ dependent on the positions of tracers (but not thermal bath particles) can be calculated according to the expression: 
\begin{equation}
<O(\{x\})>=\mathcal{N}^{-1}\int \{dPdx\} O(\{x\}) e^{-\beta (H_{tt}+H_{eff})}
\end{equation}
The effective interactions enter this formula just like another regular potential. Thus, the simplest and very common approach is to describe a dynamical system in terms of the ordinary over-damped Langevin equations, with the effective interactions included as an additional force, namely:
\begin{equation}
\begin{gathered}
\gamma \dot x_i= \sum_j^N F_0(x_i-x_j)+ \sum_j^NF_{eff}(x_i-x_j)+\eta_i(t) \label{eq:simple_Lan} \\
<\eta_i(t)\eta_j(t')>=\sigma^2\delta_{ij}\delta(t-t')
\end{gathered}
\end{equation}
Here, the friction coefficient is constant and the noise is uncorrelated. Writing the complementary Fokker-Planck equation for this system and solving it in the stationary state, one obtains the Boltzmann distribution \eqref{eq:P_Boltz}. This means that one can use \eqref{eq:simple_Lan} to obtain the averaged-out observables $<O(\{x\})>$ in the equilibrium. However, this does not guarantee that the trajectories generated by \eqref{eq:simple_Lan} are close to the actual physical trajectories, it only ensures the statistical agreement. Another problem is that in this model the average values of the observables related to the thermal bath are directly affected by the choice of uncorrelated Gaussian noise in \eqref{eq:simple_Lan}. 

The equation \eqref{eq:corr} suggests that the role of effective interactions in the dynamics, even in equilibrium, is more complicated than being simply \emph{another} potential. In the light of \eqref{eq:corr}, effective interactions might be understood as dynamically induced by the correlated behavior of thermal bath. Intuitively, one expects then that the friction and noise (which replace the deterministic tracer-bath coupling) should also reproduce these spatial correlations. Satisfying this intuition alongside the requirement of the thermodynamic consistency leads to the inclusion of SCN and SVFC. However, SCN and SVFC make the diffusion a collective phenomenon, i.e. the thermal noise affecting different particles is not independent and SVFC takes into account that one particle disturbs the environment perceived by the other particles. In Ref. \cite{bib:majka1} this manifests via the additional response forces that arise when absolute positions of particles are considered. Obviously, the thermodynamically consistent Langevin equations with SCN \cite{bib:majka1} are not perfectly equivalent to the microscopic, deterministic theory. One reason for that is e.g. the choice of the Gaussian noise and currently we cannot assess how different it is from the actual distribution of the coupling forces. However, via SCN and SVFC we are able to transfer more properties from the microscopic level into the stochastic dynamics, e.g. some microscopic self-assembly mechanisms. This is because both SCN and SVFC depend explicitly on the spatial correlation function $h(r)$, which is sensible to the thermodynamical state of the system \cite{bib:binney}. This opens the possibility of formulating a single stochastic model adequate for a few different regimes.

Another aspect of this is the non-equilibrium regime. The agreement between the microscopic theory (given by the Boltzmann distribution) and the stochastic process approach can be ensured only in equilibrium conditions. However, the Langevin dynamics also provides an insight into the non-equilibrium regime, if it is started from the non-equilibrium initial conditions. In this case, one expects that the Langevin equations of motion resemble the microscopic equations of motion closely enough, so the inference about the system evolution is allowed. From this perspective, by incorporating the spatial correlations in the noise, we construct the model which is closer to the actual physics than the non-correlated one. Thus, using SCN is possibly more accurate in the non-equilibrium regime than neglecting the correlations entirely. Obviously, since our model makes use of the equilibrium correlation function, it is still limited to the close-to-equilibrium regime. However, even under this constraint, our formalism can reveal new collective, non-equilibrium effects such as e.g. the transient attraction \cite{bib:majka1}.

\section{Multi-particle dynamics with SCN}\label{sec:multiparticle}
We will now propose the extension of our formalism to the one-dimensional, multi-particle systems ($N>2$). Let us introduce the random vector:
\begin{equation}
\vec{\xi}^T=(\xi_1,\dots,\xi_N)
\end{equation}
which components satisfy:
\begin{gather}
<\xi_i(t)>=0\\
<\xi_i(t)\xi_j(t')>=\sigma^2h(x_i-x_j)\delta(t-t')
\end{gather}
so $\vec{\xi}$ contains the correlated Gaussian variables. We shall omit the argument $t$ in the notation if it is not necessary. We can introduce the following correlation matrix:
\begin{equation}
<\vec{\xi}\vec{\xi}^T>=H
\end{equation}
where the entries are $H_{ij}=\sigma^2h(x_i-x_j)$. $H$ is a symmetric, real and positive definite matrix, so its eigen-decomposition can be written in the following form:
\begin{equation}
H=Q\Lambda^2Q^T=Q\Lambda Q^T Q\Lambda Q^T=G^2
\end{equation}
where $\Lambda$ is the diagonal matrix such that $\Lambda^2$ contains the eigenvalues of $H$ and matrix $Q$ is orthonormal. One can observe that the matrix $G=Q\Lambda Q^T=G^T$ is also symmetric, so, in fact:
\begin{equation}
H=G^2=G G^T
\end{equation}
The important fact is that we can generate the vector of correlated variables $\vec{\xi}$ from the linear combination of the uncorrelated Gaussian variables with the aid of $G$, namely:
\begin{equation}
\vec{\xi}=G\vec{\eta} \label{eq:vec_xi}
\end{equation}
where:
\begin{equation}
\begin{gathered}
\vec{\eta}^T=(\eta_1,\dots,\eta_N) \\
<\eta_i>=0 \\
<\eta_i(t)\eta_j(t')>=\delta_{ij}\delta(t-t')
\end{gathered}
\end{equation}
One can check that:
\begin{equation}
<(G\vec{\eta})(G\vec{\eta})^T>=G<\vec{\eta}\vec{\eta}^T>G^T=H
\end{equation}

From the two-particle case we know that the use of SCN requires SVFC. We also know that in absolute variables SVFC translated into the friction coefficient and the response forces \cite{bib:majka1}. Therefore, we postulate the following generalization from the two-tracer model to the case of the $N$-tracer system:
\begin{equation}
\sum_{i}^N K_{ij}\dot x_j=\sum_{j}^N F(x_i-x_j)+\xi_j \label{eq:multiparticle}
\end{equation}
where $K_{ii}$ is SVFC for the $i$-th tracer and $K_{ij\neq i}$ are the response forces. We assume that $K_{ij}$ are the functions of $x_1,\dots, x_2$, but we will omit the arguments for a more compact notation. Using \eqref{eq:vec_xi}, we can rewrite the equations of motion in the matrix form:
\begin{equation}
K\dot{\vec{x}}=\vec{F}+G\vec{\eta} \label{eq:m1}
\end{equation}
where the $i$-th component of the vector $\vec{F}$ reads:
\begin{equation}
F_i=\sum_j^NF(x_i-x_j)
\end{equation}
Our goal is to determine the entries of the matrix $K$. We begin with putting \eqref{eq:m1} into the following form:
\begin{equation}
\dot{\vec{x}}=K^{-1}GG^{-1}\vec{F}+K^{-1}G\vec{\eta} \label{eq:m2}
\end{equation}
where we assume that $K$ and $G$ are invertible. Let us now introduce the auxiliary matrix $S$:
\begin{equation}
S=K^{-1}G
\end{equation}
so our problem reads:
\begin{equation}
\dot{\vec{x}}=SG^{-1}\vec{F}+S\vec{\eta} \label{eq:m2}
\end{equation}
Applying the Stratonovich interpretation for this system, its stationary Fokker-Planck equation reads:
\begin{equation}
0=\sum_i^N \partial_{x_i}\left[ \sum_j^N S_{ij}\left(\sum_k^N\tilde g_{jk}F_kP+\frac{1}{2}\sum_k^N\partial_{x_k}(S_{kj}P) \right) \right] \label{eq:FPE1}
\end{equation}
where $\tilde g_{ij}=(G^{-1})_{ij}$ and $P_s=P(x_1,\dots,x_N)$ is the stationary probability distribution describing the system. Let us now demand that $P=P_B$, i.e. that our system tends to the Boltzmann distribution in the stationary state. Employing the definition \eqref{eq:P_Boltz} of $P_B$ in \eqref{eq:FPE1} we obtain the equation:
\begin{equation}
0=\sum_i^N \partial_{x_i}\left[ P_B\sum_j^N S_{ij}\sum_k^N\left(\tilde g_{jk}F_k +\frac{\beta}{2}F_k S_{kj}+\frac{1}{2}\partial_{x_k}S_{kj}\right) \right] \label{eq:FPE2}
\end{equation}
At this point we should specify the boundary conditions. In general, one might observe that when particles are far from each other, the correlation matrix $H$ become diagonal ($H_{ij}\to\sigma^2\delta_{ij}$) and so does $G_{ij}\to\sigma\delta_{ij}$. In this situation we also expect that $K_{ij}\to \gamma\delta_{ij}$, so, eventually, the $S_{ij}\to \frac{\sigma}{\gamma}\delta_{ij}$. However, even under these conditions we still have some significant freedom in the structure of matrix $S$. One particularly simple choice is to demand that $S$ is in fact diagonal, so $S_{ij}=S_{ii}\delta_{ij}$. One can check now that:
\begin{equation}
\begin{split}
S^2&=<(S\vec{\eta})(S\vec{\eta})^T>=(K^{-1})GG^T (K^{-1})^T\\
&=K^{-1}H(K^{-1})^T
\end{split}
\end{equation}
and for the diagonal $S$ this means the $K^{-1}$ is the transformation diagonalizing $H$, i.e. $K^{-1}=Q$.

Under this choice, the Fokker-Planck equation simplifies into:
\begin{equation}
0=\sum_i^N \partial_{x_i}\left[ P_B S_{ii}\left(\sum_k^N\tilde g_{ik}F_k +\frac{\beta}{2}F_i S_{ii}+\frac{1}{2}\partial_{x_i}S_{ii}\right) \right] \label{eq:FPE2}
\end{equation}
Taking into acocunt the boundary condition, the solution for each $S_{ii}$ reads:
 \begin{equation}
 \begin{split}
 &S_{ii}=e^{\beta\sum_{m\neq i}^NU(x_i-x_m)}\times\\
 &\left(\frac{\sigma}{\gamma}-\int_{x_i}^{+\infty}dy_i\sum_k^N\tilde g_{ik}F_k e^{-\beta\sum_{n\neq i}^NU(y_i-x_n)}\right)
 \end{split}\label{eq:S_sol}
 \end{equation}
 Note that in the product of functions $\tilde g_{ik}F_k$ the argument $x_i$ is replaced by the integration variable $y_i$.
 
Having found the matrix $S$ one obtains the matrix of SFVC and response forces via transformation $K=GS^{-1}$. One can see that certain properties from the two-particle model might transfer into the multi-particle model. In particular, it might be possible that for a certain choice of $F(x_i-x_j)$ and $h(x_i-x_j)$ there exists such set of positions $x_1^0,\dots,x_N^0$ that $S_{ii}(x_1^0,\dots,x_N^0)=0$. In this case the SVFC for the $i$-th particle would become singular and the effects analyzed in Ref. \cite{bib:majka1} might arise. However, this time we deal with the collective behavior in a sense that all particles have influence on the value of $S_{ii}$. Practically, when both $h(r)$ and $F(r)$ have finite range, only the interactions with some limited number of neighbors are significant. It is then possible to truncate the sum $\sum_k^N\tilde g_{ik}F_k$ in \eqref{eq:S_sol}, e.g. to $k=i-1, i, i+1$. This means that $S_{ii}=0$ for a single particle might be ensured by the proper choice of $x_{i-1}^0, x_{i}^0, x_{i+1}^0$ and so it could be possible to find such set of $\{x_i^0\}$ that satisfies $S_{ii}=0$ for all $i$ simultaneously. This would indicate the global change in system characteristic. However, obtaining this solution is numerically difficult and we shall not purse it in this paper.
 
Finally, we must comment that the solution \eqref{eq:S_sol} is not unique and it is possible to solve \eqref{eq:FPE2} e.g. without demanding that $S$ is diagonal. We should also denote that \eqref{eq:S_sol} leads to the non-symmetric matrix $K\neq K^T$. Whether the response forces should be reciprocal or not ($K_{ij}=K_{ji}$) should be a subject of further research. However, imposing  on \eqref{eq:FPE2} the constraints that ensure the symmetry of $K$ complicates the problem significantly and the solution in this case is also unknown. 

\section{2-particle Mori-Zwanzig model of simultaneous diffusion} \label{sec:2particleMZmodel}
The SCN-driven Langevin dynamics which we have discussed avoids the direct microscopic considerations in favor of some general considerations. For this reason it has a few ambiguities that cannot be addressed within its own framework.  The first thing is that one cannot tell whether $F_{eff}(r)$ should be present in \eqref{eq:SCN_Lan} and \eqref{eq:multiparticle} explicitly or it should manifest only indirectly, via SCN and  SVFC. In fact, it is equally possible to construct a dynamics similar to \eqref{eq:SCN_Lan}, but without the $F_{eff}(r)$ term and our approach gives no criterion to distinguish between them. Another issue is that in equation \eqref{eq:SCN_Lan} as $r\to0$ we have also $g_-(r)\to0$ and $K(r)\to0$. This means that our whatsoever over-damped theory leads to the essentially frictionless (hence inertial) dynamics for small $r$. It should be answered whether this is an artifact of the theory or a valid prediction. Finally, our dynamics incorporates SCN only, while completely abandoning the temporal correlations. One might ask then if there is a way to circumvent this limitation. We will answer these questions by undertaking a completely alternative approach, which is the multi-tracer variant of the renowned Mori-Zwanzig theory \cite{bib:zwanzig,bib:hanggi}.

In this approach we assume that $N$ tracers are coupled to the common heat-bath of $\tilde N$ oscillators. The Hamiltonian of our system is given by the equations \eqref{eq:H1}-\eqref{eq:H2}, where we specify:
\begin{equation}
V(x_i-q_n)=\frac{m w_{in}^2}{4}\left(q_n-\frac{2c_{in}}{mw_{in}^2}x_i\right)^2 \label{eq:MZint},
\end{equation}
$U_0(x_i-x_j)$ is some arbitrary interaction and $v(q_i-q_j)=0$. Adopting the most general approach we assume that $c_{in}$ and $w_{in}$ might have arbitrary values. Following the approach of Mori and Zwanzig \cite{bib:zwanzig, bib:hanggi}, we can write the equations of motion for this system:
\begin{equation}
\begin{gathered}
M\dot x_i=P_i \\
\dot P_i=-\sum_j^N\partial_{x_i}U_0(x_i-x_j)+\sum_n^{\tilde N} c_{in}\left(q_n-\frac{2c_{in}}{m w_{in}^2}x_i \right) \label{eq:main}\\
m \dot q_n=p_n \\ 
\dot p_n=-mq_n\sum_i^N\frac{w_{in}^2}{2}+\sum_i^Nc_{in}x_i
\end{gathered}
\end{equation}
Let us denote:
\begin{equation}
\omega_n^2=\sum_i^N\frac{w_{in}^2}{2}
\end{equation}
The solution for $q_n(t)$ reads \cite{bib:hanggi}:
\begin{equation}
\begin{split}
q_n(t)=&q_n(t_0)\cos(\omega_n(t-t_0))+\frac{p_n(t_0)}{m\omega_n}\sin(\omega_n(t-t_0))+\\
&+\sum_i^N\frac{c_{in}}{m\omega_n}\int_{t_0}^{t}ds\sin(\omega_n(t-s))x_i(s)
\end{split} \label{eq:q_sol}
\end{equation}
The classical move now is to apply the integration by parts in the convoluted term:
\begin{equation}
\begin{split}
&\sum_i^N\frac{c_{in}}{m\omega_n}\int_{t_0}^{t}ds\sin(\omega_n(t-t_0))x_i(s)= \\
&\sum_i^N\frac{c_{in}}{m\omega_n^2}x_i(t)-\sum_i^N\frac{c_{in}}{m\omega_n^2}x_i(t_0)\cos(\omega_n(t-t_0))-\\
&-\sum_i^N\frac{c_{in}}{m\omega_n^2}\int_{t_0}^{t}ds\cos(\omega_n(t-s))\dot x_i(s)
\end{split} \label{eq:expand}
\end{equation}
and carefully substitute $q_n(t)$ with the use of \eqref{eq:expand} into the equation of motion \eqref{eq:main}. Assigning $F_0(x_i-x_j)=-\partial_{x_i}U_0(x_i-x_j)$ we obtain the following equation:
\begin{equation}
\begin{split}
&M\ddot x_i(t)=\sum_j^NF_0(x_i-x_j)+\sum_j^NF_{eff}(x_i,x_j)-\\
&-\sum_j^N\int_{t_0}^{t}ds\mathcal{K}_{ij}(t-s)\dot x_j(s)+\xi_i(t)
\end{split}
\end{equation}
where:
\begin{gather}
F_{eff}(x_i,x_j)=-\sum_{n}^{\tilde N}\left(\frac{2c_{in}^2}{Nmw_{in}^2}x_i-\frac{c_{in}c_{jn}}{m\omega_n^2}x_j\right) \label{eq:MZ_Feff}\\
\mathcal{K}_{ij}(t-s)=\sum_n^{\tilde N}\frac{c_{in}c_{jn}}{m\omega_n^2}\cos(\omega_n(t-s)) \label{eq:MZkernel}\\
\begin{split}
&\xi_i(t)=\sum_n^{\tilde N}\frac{c_{in} p_n(t_0)}{m\omega_n}\sin(\omega_n(t-t_0))+\\
&+\sum_n^{\tilde N}c_{in}\cos(\omega_n(t-t_0))\left(q_n(t_0)- \sum_j^Nx_j(t_0) \frac{c_{jn}}{m\omega_n^2}\right)
\end{split} \label{eq:xi_MZ}
\end{gather}
The above theory differs from the single-particle Generalized Langevin Equation \cite{bib:zwanzig,bib:hanggi} in a few significant aspects. First, let us look at the friction term $\sum_j\int_{t_0}^tds\mathcal{K}_{ij}(t-s)\dot x_j(s)$. While it has the usual form of the convoluted memory kernel, it has the contributions from every tracer. This closely resembles the response forces appearing in Ref. \cite{bib:majka1}, so the dissipation of energy by one tracer is affected by the others. Another issue is the stochastic force term $\xi_i(t)$, which also contains the contributions from all tracers.

Let us now restrict to the $N=2$ system and assume that the $c_{in}=c_{jn}$ for every pair of $i$ and $j$. When we switch to the relative distance $r=x_2-x_1$ and the center of the mass $X=(x_1+x_2)/2$ variables, the equations of motion read:
\begin{gather}
M\ddot r=2F_0(r)+2F_{eff}(X-\frac{r}{2},X+\frac{r}{2}) \label{eq:delta_x}\\
\begin{split}
\frac{M}{2}\ddot X=&\xi(t)-2\int_{t_0}^{t}dsK(t-s)\dot X(s)
\end{split}
\end{gather}
One can instantly notice that the dynamics of $r$ becomes purely deterministic, i.e. equation \eqref{eq:delta_x} is affected only by $F_{eff}$, but not by $\xi_i(t)$ or the memory kernel. All the 'stochasticity' is poured into the dynamics of the mass center. However, this is possible due to the symmetric choice of $c_{in}$. Yet another issue is the emergence of the additional inter-particle force \eqref{eq:MZ_Feff}. This result might be linked to the effective interactions theory. Inasmuch $\mathcal{H}_{eff}$ given by \eqref{eq:Ueff} traces out $q_n$ and $p_n$ from the partition function, the Mori-Zwanzig approach renormalizes the equations of motion. It is then welcome that some additional force $F_{eff}(r)$ emerges in \eqref{eq:delta_x}. However, since we do not assume the equilibrium condition in Mori-Zwanzig approach, this additional force might be seen as a non-equilibrium effective interaction. However, the force \eqref{eq:MZ_Feff} is generally non-local, i.e. it depends on the absolute positions of particles. This effect is not present provided that the symmetry in coupling (for both $c_{in}=c_{jn}$ and $w_{in}=w_{jn}$, $i\neq j$) is assumed. In this case $F_{eff}(r)=-\sum_{n}^{\tilde N}\frac{c_n^2}{mw_n^2}r$, which is local. 

We now get back to the general $N$-tracer case and examine what kind of spatial correlations in noise are predicted by the Mori-Zwanzig model. Typically, we shall assume that the initial  positions and momenta satisfy the equilibrium Boltzmann distribution \cite{bib:hanggi}. From \eqref{eq:xi_MZ}, one can see that $\xi_i(t)$ is the function of $x_i(t_0)$, so we will denote $\xi_i(t)=\xi(x_i(t_0),t)$ for the $i$-th particle. We want to calculate the covariance function $<\xi(x_i(t_0),t)\xi(x_j(t_0)+r,s)>$, where the averages are taken over the every degree of freedom in $t=t_0$. From the structure of \eqref{eq:xi_MZ} we conclude that $\xi(x_j(t_0)+r,t)$ simply reads:
\begin{equation}
\xi(x_j(t_0)+r,t)=\xi(x_j(t_0),t)-r\sum_n^{\tilde N}\frac{c_{jn}^2}{m\omega_n}\cos(\omega_n(t-t_0))
\end{equation}
Thus, the correlation function reduces to:
\begin{equation}
\begin{split}
&<\xi(x_i(t_0),t)\xi(x_j(t_0)+r,s)>=\\
&<\xi_i(t)\xi_j(s)>-r<\xi_j(s)>\sum_n^{\tilde N}\frac{c_{jn}^2}{m\omega_n^2}\cos(\omega_n(s-t_0))
\end{split}
\end{equation}
In order to quickly calculate $<\xi(x_j(t_0),t)>$ it is feasible to rearrange the interaction terms that enter the Boltzmann distribution:
\begin{equation}
\begin{split}
&\sum_{i}^N\sum_n^{\tilde N}V(x_i-q_n)=\sum_{n}^{\tilde N}\frac{m\omega_n^2}{2}\left(q_n-\sum_i^N\frac{c_{in}}{m\omega_n^2}x_i \right)^2-\\
&-\sum_n^N\frac{1}{m\omega_n^2}\left(\sum_i^Nc_{in}x_i \right)^2+\sum_i^N\sum_n^{\tilde N}\frac{c_{in}}{mw_{in}^2}x_i^2
\end{split}
\end{equation}
This first sum instantly matches the structure of ${<\xi(x_j(t_0),t)>}$ given by \eqref{eq:xi_MZ}, so the integrations over $p_n(t_0)$ and $q_n(t_0)$ in $<\xi(x_j(t_0),t)>$ lead to:
\begin{equation}
<\xi(x_j(t_0),t)>=0
\end{equation}
In fact, the above result is expected for the stochastic force. We can also calculate $<\xi_i(t)\xi_j(s)>$ to conclude:
\begin{equation}
<\xi(x_i(t_0),t)\xi(x_j(t_0)+r,s)>=\frac{1}{\beta}\sum_n^{\tilde N}\frac{c_{in}c_{jn}}{m\omega_n^2}\cos(\omega_n(t-s)) \label{eq:MZ_corr_fun}
\end{equation}
This recovers the celebrated relation between the temporal correlations in noise and friction memory kernel \cite{bib:zwanzig, bib:hanggi, bib:kou}. In the purely spatial context the stochastic forces prove extremely strongly spatially correlated, i.e. for $t=s$:
\begin{equation}
<\xi(x_i(t_0),t)\xi(x_j(t_0)+r,t)>=\frac{1}{\beta}\sum_n^{\tilde N}\frac{c_{in}c_{jn}}{m\omega_n^2}
\end{equation}
which is constant in entire space. In fact, on average, all tracers are moved by the thermal bath as one entity.

\section{Mori-Zwanzig model vs. SCN-driven Langevin dynamics}\label{sec:comparison}
We will now summarize the properties of Mori-Zwanzig approach and compare it to the SCN-based dynamics of the multi-tracer diffusion. The Mori-Zwanzig approach is fully microscopically consistent and, at least formally, it describes the spatio-temporal aspects of simultaneous diffusion. Yet, the Mori-Zwanzig model is limited to the linear tracer-bath coupling and neglects the bath-bath interactions, unlike our model, which is not restricted in these aspects. Both models agree in that the dissipative term for a single tracer can be decomposed into the proper friction term and the sum of response forces, which are the contributions from the other tracers. This supports the idea of the collective diffusion.

One question regarding our SCN-driven dynamics is whether or not the effective forces should be explicitly included in the Langevin equation \eqref{eq:SCN_Lan}. The observation retrieved from the Mori-Zwanzig model \eqref{eq:delta_x} is that the effective force emerges in the renormalized equation of motion, thus justifying the former option. Although the effective interactions predicted by the Mori-Zwanzig model are limited only to the relatively simple form \eqref{eq:MZ_Feff}, it is also remarkable that they can be both equilibrium or non-equilibrium. This is also in contrast with Ref. \cite{bib:majka1}, where we are limited to the equilibrium effective interactions. 

Another problem is the $K(r)\to 0$ and $g_-(r)\to0$ behavior as $r\to0$ in \eqref{eq:SCN_Lan}, which is the limiting case for our over-dapmed dynamics. Our theory predicts that the influence of both friction and noise should vanish. The Mori-Zwanzig model shows that, indeed, the dynamics of $r$ can become purely deterministic provided that there is a symmetry in the coefficients $c_{in}=c_{jn}$ for $i\neq j$ in \eqref{eq:MZint}. In this case the equation of motion for $r$ reads \eqref{eq:delta_x}, which has no stochastic contribution. 

The problem of symmetry $c_{in}=c_{jn}$ for $i\neq j$ is related to the interpretation of the coupling potential \eqref{eq:MZint}. If we perceive it as a quadratic expansion of some otherwise non-linear function there is the question of whether we expand this nonlinear $V(x_i-q_n)$ about the same position for every tracer (e.g. $V(x_i-q_n)\simeq V(0)+V'(0)(x_i-q)+\frac{1}{2}V''(0)(x_i-q)^2$) or not. In the former case the symmetry is naturally ensured, but not in the latter. This is also partially related to the range of applicability of Mori-Zwanzig model. On the one hand, it can be constructed wherever the linear approximation applies to the coupling forces. This usually requires a local energy minimum. On the other hand, the Mori-Zwanzig model predicts the constant spatial correlation function of stochastic forces, but the realistic expectation (confirmed by the microscopic relation \eqref{eq:corr}) is a function which decays with distance. Thus, one can see the Mori-Zwanzig model as a limit of strong correlations and identify it with the short-distance limit of SCN-based theory. At the same time, this latter theory seems applicable as long as the equilibrium correlation function and the Gaussian SCN are adequate approximations. 

Yet another important aspect of the multi-tracer Mori-Zwanzig model is the predicted spatio-temporal correlation function. This function shows that the entire group of tracers within some range can move in a highly coordinated manner. This resembles the behavior of glasses \cite{bib:glass0,bib:glass1,bib:glass2,bib:glass3,bib:glass4}. It is also well known that the temporal correlations in the motion of a single tracer can result in the subdiffusive dynamics, i.e. its mean square displacement (MSD) grows like $\propto t^\alpha$ (where $t$ - time, $0<\alpha<1$). The single-tracer Generalized Langevi Equations (in fact: the Mori-Zwanzig theory) are commonly used to model this effect (e.g. \cite{bib:kou, bib:goychuk}). The multi-particle Mori-Zwanzig model extrapolates these predictions on the group of tracers, provided that the constants $c_{in}$ are chosen so the memory kernel \eqref{eq:MZkernel} also takes the  form of the power-law \cite{bib:kou, bib:goychuk}. On the other hand, while the subdiffusive dynamics is observed in the dense colloids, it is claimed to manifest via the logarithmic growth of MSD ($\propto\ln t/\tau$) \cite{bib:subglass1,bib:subglass2}. This impose an intriguing problem. On the one hand, the logarithmic MSD is usually modeled as a result of the non-Gaussian noise (e.g. \cite{bib:chechkin}), not the memory effect. On the other hand, the logarithmic and fractional power-law functions can mimic each other even over several orders of magnitude, when properly adjusted. Supposing that the actual experimental dependence is, in fact, of the power-law type, the Mori-Zwanzig model could be readily applied to reproduce these data. However, this would require a change in their current interpretation.

Finally, let us comment on the limitations of the Mori-Zwanzig approach. The main restriction is the linearity in coupling, but, unfortunately it is also its essential ingredient. Linearity results in the somewhat over-simplified form of the effective interaction \eqref{eq:MZ_Feff} and in the constant spatial correlation function.  This last result is particularly unrealistic, especially for huge distance $r$ between the tracers. These impediments could be partially remedied by the coupling force of the quasi-non-linear form e.g. $(q_n-x_i)G(x_i)$, for which it is also possible to obtain some analytical results \cite{bib:hanggi}. However, such coupling is not invariant under the global translations (i.e. $((q_n+\epsilon)-(x+\epsilon))G(x_i+\epsilon)\neq(q_n-x_i)G(x_i)$), so this solution is also dissatisfying. All of these problems show that the Mori-Zwanzig approach can serve only as the auxiliary short-distance model and a more versatile workaround is necessary, such as e.g. our SCN-driven Langevin dynamics. Obviously, our theory is restricted to the SCN-dominated systems as it cannot handle the temporal correlations. For this reason, the Mori-Zwazig model suggest that a similar approach, but covering the non-linear coupling could be the ultimate solution to the problem of spatio-temporal correlations in dynamics.

\section{Summary}
The theory of diffusion in colloids has gone a long way from its initial formulations up to the present days. Its significance only grew as it has been involved in the understanding of the biological and self-organizing systems. In this paper we have attempted to review and summarize the growing field of research on the SCN-related phenomena. We have also expounded the idea of collectivity in diffusion and provided two complementary models that reflect this idea. The collective perspective expands the importance of the diffusion theory to the field of molecular self-assembly and sheds a new light on the stochastic dynamics of transient, non-equilibrium states. The theory of thermodynamically consistent SCN-driven Langevin equations \cite{bib:majka1} and the multi-particle Mori-Zwanzig model provided in this article mount to a coherent picture, establishing the framework for collective diffusion. However they must be considered as some intermediate steps towards the even more comprehensive theory of spatio-temporal correlations which could possibly deal with the non-equilibrium states without resorting to the equilibrium or linear approximations. 

\begin{acknowledgments}
The authors gratefully acknowledges the National Science Center, Poland for the grant support (2014/13/B/ST2/02014).
\end{acknowledgments}

\end{document}